\documentstyle[epsfig]{aipproc}

\begin{document}
\title{Revelations of the U(1)-Extended Supersymmetric Standard Model}

\author{E. Keith and Ernest Ma}
\address{Department of Physics,
University of California\\
Riverside, CA 92521, USA}

\maketitle

\begin{abstract}
If an extra supersymmetric U(1) gauge factor exists at the TeV 
energy scale, which is then broken together with the supersymmetry, 
there will be several interesting and important phenomenological consequences, 
not only at the TeV scale, but also at the 100 GeV scale.  For one, the 
generic two-doublet Higgs structure will involve 3 additional parameters  
beyond that of the Minimal Supersymmetric Standard Model (MSSM), thereby 
raising the upper bound on the mass of the lighter of the two neutral Higgs 
scalars.  For another. the supersymmetric scalar quarks and leptons 
receive new contributions to their masses from the spontaneous breaking 
of this extra U(1).  Assuming a grand-unified $E_6$ gauge symmetry and 
universal soft supersymmetry-breaking terms at the grand-unification 
energy scale, we find solutions relating the U(1) breaking scale and 
the ratio of the vacuum expectation values of the two electroweak 
Higgs doublets.
\end{abstract}

\section*{Two-Doublet Higgs Structure}

The possible existence of a supersymmetric U(1) gauge factor\cite{1,2,3,4} 
which is broken together with the supersymmetry at the TeV scale will 
affect the physics at the electroweak scale, through tree-level 
nondecoupling effects in the Higgs sector.\cite{5}  Specific examples 
have been given previously.\cite{6,7,8}  Here we consider the most general 
case.\cite{9}  We assume the Higgs sector to consist of two doublets and 
a singlet.  Under $SU(3)_C \times SU(2)_L \times U(1)_Y \times U(1)_X$,
\begin{eqnarray}
\tilde \Phi_1 &=& \left( \begin{array} {c} \bar \phi_1^0 \\ -\phi_1^- 
\end{array} \right) \sim (1,2,-{1 \over 2}; -a), \\ 
\Phi_2 &=& \left( \begin{array} {c} \phi_2^+ \\ \phi_2^0 \end{array} 
\right) \sim (1,2, {1 \over 2}; -1 + a), \\ 
\chi &=& \chi^0 \sim (1,1,0;1),
\end{eqnarray}
where each last entry is the arbitrary assignment of that scalar multiplet 
under the extra $U(1)_X$ with coupling $g_x$ such that a term in the 
superpotential exists linking all three superfields with coupling $f$. 
Consequently, the quartic terms of the scalar potential are given by
\begin{eqnarray}
V &=& f^2 [ (\Phi_1^\dagger \Phi_2)(\Phi_2^\dagger \Phi_1) + (\Phi_1^\dagger 
\Phi_1 + \Phi_2^\dagger \Phi_2) \bar \chi \chi ] \nonumber \\ 
&+& {1 \over 8} g_2^2 [(\Phi_1^\dagger \Phi_1)^2 + (\Phi_2^\dagger \Phi_2)^2 + 
2 (\Phi_1^\dagger \Phi_1)(\Phi_2^\dagger \Phi_2) - 4 (\Phi_1^\dagger \Phi_2) 
(\Phi_2^\dagger \Phi_1)] \nonumber \\ 
&+& {1 \over 8} g_1^2 [-\Phi_1^\dagger \Phi_1 + \Phi_2^\dagger \Phi_2]^2 
+ {1 \over 2} 
g_x^2 [-a \Phi_1^\dagger \Phi_1 - (1-a) \Phi_2^\dagger \Phi_2 + \bar \chi 
\chi]^2,
\end{eqnarray}
Let $\langle \chi \rangle = u$, then $\sqrt 2 Re \chi$ is a physical scalar 
boson with $m^2 = 2 g_x^2 u^2$, and the $(\Phi_1^\dagger \Phi_1) \sqrt 2 Re 
\chi$ coupling is $\sqrt 2 u (f^2 - g_x^2 a)$.  Hence the effective 
$(\Phi_1^\dagger \Phi_1)^2$ coupling $\lambda_1$ is given by
\begin{equation}
\lambda_1 = {1 \over 4}(g_1^2 + g_2^2) + g_x^2 a^2 - {{2(f^2 - g_x^2 a)^2} 
\over {2 g_x^2}} = {1 \over 4} (g_1^2 + g_2^2) + 2 a f^2 - {f^4 \over g_x^2}.
\end{equation}
Similarly,
\begin{eqnarray}
\lambda_2 &=& {1 \over 4} (g_1^2 + g_2^2) + 2(1-a) f^2 - {f^4 \over g_x^2}, \\ 
\lambda_3 &=& -{1 \over 4} g_1^2 + {1 \over 4} g_2^2 + f^2 - {f^4 \over 
g_x^2}, \\ \lambda_4 &=& -{1 \over 2} g_2^2 + f^2,
\end{eqnarray}
where the two-doublet Higgs potential has the generic form
\begin{eqnarray}
V &=& m_1^2 \Phi_1^\dagger \Phi_1 + m_2^2 \Phi_2^\dagger \Phi_2 + m_{12}^2 
(\Phi_1^\dagger \Phi_2 + \Phi_2^\dagger \Phi_1) + {1 \over 2} \lambda_1 
(\Phi_1^\dagger \Phi_1)^2 \nonumber \\ &~& + {1 \over 2} \lambda_2 
(\Phi_2^\dagger 
\Phi_2)^2 + \lambda_3 (\Phi_1^\dagger \Phi_1) (\Phi_2^\dagger \Phi_2) + 
\lambda_4 (\Phi_1^\dagger \Phi_2)(\Phi_2^\dagger \Phi_1).
\end{eqnarray}
In the above, we have assumed that the soft supersymmetry-breaking term 
$f A_f \Phi_1^\dagger \Phi_2 \chi + h.c.$ (from which we obtain $m_{12}^2 
= f A_f u$) is small, otherwise the electroweak Higgs sector reduces 
effectively to just one light doublet.  Note also that the Minimal 
Supersymmetric Standard Model (MSSM) is recovered in the limit $f=0$, 
independent of $g_x^2$ and $a$.

Let $\langle \phi_{1,2}^0 \rangle \equiv v_{1,2}$, $\tan \beta \equiv 
v_2/v_1$, and $v^2 \equiv v_1^2 + v_2^2$, then the above $V$ has an upper 
bound on the lighter of the two neutral scalar bosons given by
\begin{equation}
(m_h^2)_{max} = 2 v^2 [\lambda_1 \cos^4 \beta + \lambda_2 \sin^4 \beta + 
2 (\lambda_3 + \lambda_4) \sin^2 \beta \cos^2 \beta] + \epsilon,
\end{equation}
where we have added the radiative correction due to the $t$ quark and its 
supersymmetric scalar partners, {\it i.e.}
\begin{equation}
\epsilon = {{3 g_2^2 m_t^4} \over {8 \pi^2 M_W^2}} \ln \left( 1 + {\tilde m^2 
\over m_t^2} \right).
\end{equation}
The existence of a supersymmetric U(1) gauge factor thus implies
\begin{equation}
(m_h^2)_{max} = M_Z^2 \cos^2 2 \beta + \epsilon + {f^2 \over {\sqrt 2 G_F}} 
\left[ A - {f^2 \over g_x^2} \right],
\end{equation}
where
\begin{equation}
A = {3\over 2} + (2a-1) \cos 2 \beta - {1 \over 2} \cos^2 2 \beta.
\end{equation}
If $A > 0$, then the MSSM bound of 128 GeV can be exceeded.  However, $f^2$ 
is still constrained because $V$ of Eq.~(9) has to be bounded from below.  
For a given $g_x^2$, we can vary $f^2$, $a$, and $\cos 2 \beta$ to find the 
largest $m_h$, which increases with increasing $g_x^2$.  We find that for 
$g_x^2 = 0.5$, $m_h < 190$ GeV.  On the other hand, for specific 
models\cite{3,4,6} where $a$ and $g_x^2$ are known, the largest $m_h$ is 
only between 140 and 145 GeV.\cite{9}

\section*{Supersymmetric Scalar Masses}

Let us specify the extra U(1) to be that which comes from the reduction of 
$E_6$ to the standard $SU(3)_C \times SU(2)_L \times U(1)_Y$.  In the 
chain $E_6 \rightarrow SO(10) \times U(1)_\psi$, then $SO(10) \rightarrow 
SU(5) \times U(1)_\chi$, we assume that
\begin{equation}
U(1)_\psi \times U(1)_\chi \rightarrow U(1)_\alpha,
\end{equation}
which then survives down to the TeV energy scale.  Under $E_6 \rightarrow 
SU(5) \times U(1)_\psi \times U(1)_\chi$, the fundamental {\bf 27} 
representation of $E_6$ is decomposed as follows:
\begin{eqnarray}
{\bf 27} &=& (10; 1, -1) + (5^*; 1, 3) + (1; 1, -5) \nonumber \\ 
&+& (5; -2, 2) + (5^*; -2, -2) + (1; 4, 0),
\end{eqnarray}
where $2 \sqrt 6 Q_\psi$ and $2 \sqrt {10} Q_\chi$ are denoted.  The usual 
quarks and leptons belong to $(10; 1, -1)$ and $(5^*; 1, 3)$, whereas the 
two Higgs doublets are in $(5; -2, 2)$ and $(5^*; -2, -2)$.  The $(1;1,-5)$ 
is identifiable with the right-handed neutrino $N$ and the $(1;4,0)$ is 
the singlet $S$ whose scalar component is $\chi$.  Let 
\begin{equation}
Q_\alpha = Q_\psi \cos \alpha - Q_\chi \sin \alpha,
\end{equation}
then the so-called $\eta$-model\cite{1,3} is obtained with $\tan \alpha = 
\sqrt {3/5}$ and we have
\begin{equation}
{\bf 27} = (10;2) + (5^*; -1) + (1;5) + (5; -4) + (5^*; -1) + (1;5),
\end{equation}
where $2 \sqrt {15} Q_\eta$ is denoted; and the $N$-model\cite{4,8} is 
obtained with $\tan \alpha = -1/\sqrt {15}$ resulting in
\begin{equation}
{\bf 27} = (10;1) + (5^*;2) + (1;0) + (5;-2) + (5^*;-3) + (1;5),
\end{equation}
where $2 \sqrt {10} Q_N$ is denoted.

Consider now the masses of the supersymmetric scalar partners of the quarks 
and leptons:
\begin{equation}
m_B^2 = m_0^2 + m_R^2 + m_F^2 + m_D^2,
\end{equation}
where $m_0$ is a universal soft supersymmetry breaking mass at the 
grand-unification scale, $m_R^2$ is a correction generated by the 
renormalization-group equations running from the grand-unification 
scale down to the TeV scale, $m_F$ is the explicit mass of the fermion 
partner, and $m_D^2$ is a term induced by gauge-boson masses.  In the MSSM, 
$m_D^2$ is of order $M_Z^2$ and does not change $m_B$ significantly.  In the 
$U(1)_\alpha$-extended model, $m_D^2$ is of order $M_{Z'}^2 = (4/3) 
\cos^2 \alpha g_\alpha^2 u^2$ and will affect 
$m_B$ in a nontrivial way.  Specifically,
\begin{eqnarray}
\Delta m_D^2 (10; 1, -1) &=& {1 \over 8} M_{Z'}^2 \left( 1 + \sqrt {3 \over 5} 
\tan \alpha \right), \\ \Delta m_D^2 (5^*; 1, 3) &=& {1 \over 8} M_{Z'}^2 
\left( 1 - 3 \sqrt {3 \over 5} \tan \alpha \right), \\ \Delta m_D^2 
(1; 1, -5) &=& {1 \over 8} M_{Z'}^2 ( 1 + \sqrt {15} \tan \alpha ), \\ 
\Delta m_D^2 (5; -2, 2) &=& -{1 \over 4} M_{Z'}^2 \left( 1 + \sqrt {3 \over 5} 
\tan \alpha \right), \\ \Delta m_D^2 (5^*; -2, -2) &=& -{1 \over 4} M_{Z'}^2 
\left( 1 - \sqrt {3 \over 5} \tan \alpha \right), \\ \Delta m_D^2 (1; 4, 0) 
&=& {1 \over 2} M_{Z'}^2.
\end{eqnarray}
This will have important consequences on the experimental search of 
supersymmetric particles.  In fact, depending on $m_F$, it is possible for 
exotic scalars to be lighter than the usual scalar quarks and leptons.\cite{10} 
It may explain why a scalar ``leptoquark" can be as light as 200 GeV, 
as a possible interpretation of the recent HERA data.\cite{11}

\section*{U(1) and Electroweak Breaking}

Another important outcome of Eq.~(19) is that the $U(1)_\alpha$ and 
electroweak symmetry breakings are related\cite{9}.  To see this, go back to 
the two-doublet Higgs potential $V$ of Eq.~(9).  Using Eqs.~(5) to (8), we 
can express the parameters $m_{12}^2$, $m_1^2$, and $m_2^2$ in terms of the 
mass of the pseudoscalar boson, $m_A$, and $\tan \beta$.
\begin{eqnarray}
m_{12}^2 &=& -m_A^2 \sin \beta \cos \beta, \\ m_1^2 &=& m_A^2 \sin^2 \beta 
- {1 \over 2} M_Z^2 \cos 2 \beta \nonumber \\ &~& - {{2 f^2} \over g_Z^2} 
M_Z^2 \left[ 2 \sin^2 \beta + \left( 1 - \sqrt {3 \over 5} \tan \alpha \right) 
\cos^2 \beta - {{3 f^2} \over {2 \cos^2 \alpha ~g_\alpha^2}} \right], \\ 
m_2^2 &=& m_A^2 \cos^2 \beta + {1 \over 2} M_Z^2 \cos 2 \beta \nonumber \\ 
&~& - {{2 f^2} \over g_Z^2} M_Z^2 \left[ 2 \cos^2 \beta + \left( 1 + 
\sqrt {3 \over 5} \tan \alpha \right) \sin^2 \beta - {{3 f^2} \over {2 
\cos^2 \alpha ~g_\alpha^2}} \right].
\end{eqnarray}
On the other hand, using Eq.~(19), we have
\begin{eqnarray}
m_{12}^2 &=& f A_f u, \\ m_1^2 &=& m_0^2 + m_R^2 (\tilde g, f) + f^2 u^2 - 
{1 \over 4} \left( 1 - \sqrt {3 \over 5} \tan \alpha \right) M_{Z'}^2, \\ 
m_2^2 &=& m_0^2 + m_R^2 (\tilde g, f) + f^2 u^2 - {1 \over 4} \left( 1 + 
\sqrt {3 \over 5} \tan \alpha \right) M_{Z'}^2 + m_R^2 (\lambda_t),
\end{eqnarray}
where $f A_f$ is the coupling of the soft supersymmetry-breaking 
$\tilde \Phi_1 \Phi_2 \chi$ scalar term, $\tilde g$ is 
the gluino, and $\lambda_t$ is the Yukawa coupling of $\Phi_2$ to the $t$ 
quark.  Matching Eqs.~(26) to (28) with Eqs.~(29) to (31) allows us to 
determine $u$ and $\tan \beta$ as a function of $f$ for a given set of 
parameters at the grand-unification scale.

In the MSSM assuming Eq.~(19),
\begin{equation}
m_1^2 - m_2^2 = - m_R^2 (\lambda_t) = - (m_A^2 + M_Z^2) \cos 2 \beta.
\end{equation}
Since $m_R^2 (\lambda_t) < 0$, we must have $\tan \beta > 1$.  In the 
$U(1)_\alpha$-extended model, because of the extra D-term contribution, 
$\tan \beta < 1$ becomes possible.  Another consequence is that because 
of Eq.~(20), a light scalar $t$ quark is not possible unless $\tan \alpha 
< -\sqrt {5/3}$.

To obtain the spontaneous breaking of $U(1)_\alpha$, {\it i.e.} $\langle 
\chi \rangle = u$, we need $m_\chi^2$ to be negative.  This may be achieved 
radiatively in analogy to electroweak symmetry breaking by having a relevant 
large Yukawa coupling drive $m_\chi^2$ from its universal positive value 
at the grand-unification energy scale to a negative one at the TeV scale. 
Consider the superpotential
\begin{equation}
W = f H_1 H_2 S + f' h h^c S + \lambda_t H_2 Q_3 t^c,
\end{equation}
where $h$ and $h^c$ are the exotic color triplets belonging to the 
$(5; -2, 2)$ and $(5^*; -2, -2)$ representations, as well as the corresponding 
trilinear soft supersymmetry-breaking terms in the scalar potential:
\begin{equation}
V_{soft} = f A_f \Phi_1^\dagger \Phi_2 \chi + f' A_{f'} \tilde h \tilde h^c 
\chi + \lambda_t A_t \Phi_2 \tilde Q_3 \tilde t^c + h.c.,
\end{equation}
along with all the soft supersymmetry breaking scalar masses.  We have 
verified that for $f'$ of order $\lambda_t$, it is indeed possible to break 
$U(1)_\alpha$ at the TeV scale.\cite{2,12}  However, it is highly nontrivial 
to find solutions which can satisfy Eqs.~(26) to (31) simultaneously.  
A typical solution has $f=0.345$, $m_{\tilde g} = 300$ GeV, $m_0 = A_0 = 950$ 
GeV, for which $|u| \sim 2$ TeV and $\tan \beta \sim 3.5$.  That would predict 
$M_{Z'} \sim 1$ TeV and the $h$-$h^c$ fermion mass to be about 2 TeV. 
Details are given in Ref.~[9].

\section*{Conclusions}

(1) Supersymmetric $U(1)_\alpha$ from $E_6$ is a good possiblity at the 
TeV scale. (2) The two-doublet Higgs structure at around 100 GeV will be 
different from that of the MSSM. (3) Supersymmetric scalar masses depend 
crucially on $U(1)_\alpha$. (4) The $U(1)_\alpha$ breaking scale and 
$\tan \beta$ are closely related.

\section*{Acknowledgments}

The presenter of this talk (E.M.) thanks Per Osland and Gerald Eigen for 
their great hospitality and a stimulating Beyond the Standard Model V 
meeting. This work was supported in part by the U. S. 
Department of Energy under Grant No. DE-FG03-94ER40837.

\end{document}